%% file: ABiggs.tex
\documentclass{evn2004}
\usepackage{txfonts}
\usepackage{graphicx}
\begin{document}
   \title{EVN observations of the lens system B0128+437}

   \author{A.~D. Biggs}

   \institute{Joint Institute for VLBI in Europe, Postbus 2, 7990 AA Dwingeloo, The Netherlands}

   \abstract{Previous VLBI observations of the lens system CLASS
   B0128+437 showed that the lensed source had a triple sub-component
   structure. However, one of the images looked very different to the
   others and it has been suggested that this is due to scatter-broadening
   in the ISM of the lensing galaxy. We present a new 21-cm EVN map of this
   system, the low frequency compared to previous epochs chosen as the
   scattering should be more pronounced at this frequency. Analysis of the
   new map is at an early stage, but the area of scattering in the affected
   image is extremely obvious. Other points of interest that we seek to
   explain are the very large sizes of the images (in conflict with the
   observed spectral turnover at $\sim$1~GHz) and the different position
   angle of one of the images at different frequencies.}

   \maketitle
%

\section{Introduction}

Whilst gravitational lenses attract most attention for their usefulness
in determining cosmological parameters ($H_0$, $\Omega_m$,
$\Omega_{\lambda}$) and the mass and mass profiles of lensing galaxies
(see Kochanek, Schneider \& Wambsganss \cite{kochanek04} for a recent
review), usually receiving less attention is the way in which they
can be used as probes of the interstellar medium (ISM) of high-redshift
galaxies. The galaxies in question are of course the lensing galaxies,
the deflected light from the background lensed quasar passing through
and illuminating their ISMs. The ISM can leave its imprint on the lensed
light in a number of ways e.g., differential extinction, Faraday rotation
(and depolarisation), absorption lines and scatter-broadening. From,
usually multi-frequency, observations of these phenomena it is possible
to constrain properties of the ISM, such as the magnetic field strength
and electron density, as well as the scales over which these vary. In
this way, gravitational lenses are unique probes of the astrophysics of
the high-redshift universe.

Scatter-broadening has been claimed in a number of lens systems e.g.,
PKS~1830-211 (Jones et al., \cite{jones96}, Guirado et al.,
\cite{guirado99}), JVAS B0218+357 (Biggs et al., \cite{biggs03}),
CLASS B1933+503 (Marlow et al., \cite{marlow99}) and PMN~J0134-0931
(Winn, Rusin \& Kochanek \cite{winn04}). The claims are based on
different criteria in each case. The ``classic'' manner in which to
diagnose scatter-broadening is to measure the sizes of components,
these being expected to vary according to the wavelength squared; this
has been done in the case of PKS~1830-211. For B0218+357, maps of the
two images made at 8.4~GHz with global VLBI were back-projected into
the source plane using a lens model where one image was found to be
much smoother than the other. Whilst the four images of the radio core
in B1933+503 are clearly detected with MERLIN and the VLA, two are
absent in a VLBA image and must therefore have had their surface
brightness reduced separate to the lensing process.

\begin{figure*}
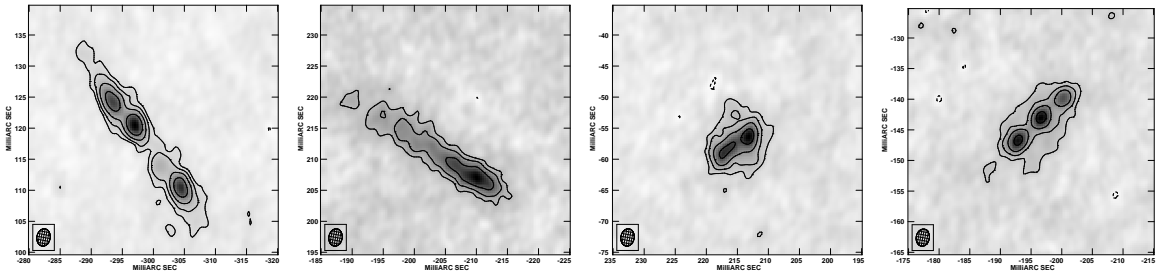

\centering
\includegraphics[scale=0.2]{ABiggs_fig1a.ps}
\includegraphics[scale=0.2]{ABiggs_fig1b.ps}
\includegraphics[scale=0.2]{ABiggs_fig1c.ps}
\includegraphics[scale=0.2]{ABiggs_fig1d.ps}
\caption{VLBA 5-GHz maps of the four images of CLASS B0128+437.
Left to right: A, B, C and D. Three sub-components are visible in
each image except for B. The two easterly sub-components in image C are
blended due to the shortness of the jet in this image, but are resolved
in a uniformly-weighted image. The restoring beam has a FWHM of $2.8 \times
2.2$~mas$^2$ at a position angle of $-10$\fdg4.\label{vlbamaps}}
\end{figure*}

Another lens system where scatter-broadening has been suggested
is CLASS~B0128+437 (Phillips et al., \cite{phillips00}). This
is a four-image lens system, the lensed source being a $z=3.12$
quasar and the maximum separation of the images only 540~mas. Originally
identified as a lens from MERLIN observations, subsequent follow-up with
the VLBA at 5~GHz (Biggs et al. \cite{biggs04}) resolved each image into
a jet with three prominent sub-components (Fig.~\ref{vlbamaps}). The
latter, however, were only present in three of the four images! Comparison
between maps made with uniform and natural weighting indicated that the
image with the missing sub-components was extended over larger
angular scales than the other images, in conflict with what would be
expected if lensing alone was determining the relative sizes of each image.
The most natural explanation for this is that image B is being
scatter-broadened in the ISM of the lensing galaxy.

In this paper we present a new map of B0128+437 made with the EVN at a
wavelength of 21~cm. The motivation for this experiment was to
investigate the source structure at an additional frequency, 8.4, 5
and 2.3~GHz observations having already been carried out with the VLBA
(including Effelsberg in S/X band). At the new lowest frequency the
scattering would be greatest and the images both brighter (the source GPS
spectrum peaks at $\sim$1~GHz) and larger. The increase in source size with
decreasing frequency was clear from the previous observations and would
give additional constraints on our models of the lensing mass distribution.


\section{The Data}

CLASS B0128+437 was observed with nine antennas of the EVN on
2004 February 18. The participating antennas were Effelsberg, Jodrell
Bank (Lovell), Medicina, Noto, Westerbork (tied array), Cambridge,
Torun, Onsala and Urumqi. The Cambridge antenna was included as it
was planned to combine these data with previous 21-cm MERLIN data.
Enhanced scattering at the low frequency might have rendered much
of the flux in the scattered image undetectable with EVN baselines
alone. Ultimately, practically all of the single-dish flux density
was recovered in the EVN map and so no combination with the MERLIN data has
taken place. The lens data were phase-referenced to a nearby compact
calibrator ($\Delta\theta < 1^{\circ}$) as even at its brightest the total
lens system flux density is only about 130~mJy. Correlation took place
at the Joint Institute for VLBI in Europe (JIVE) using the EVN MkIV Data
Processor, producing 16 spectral channels in each of the four 8~MHz 
dual-polarisation bands. The correlator averaging time was 2~s.

\begin{figure*}
\centering
\includegraphics[scale=0.75]{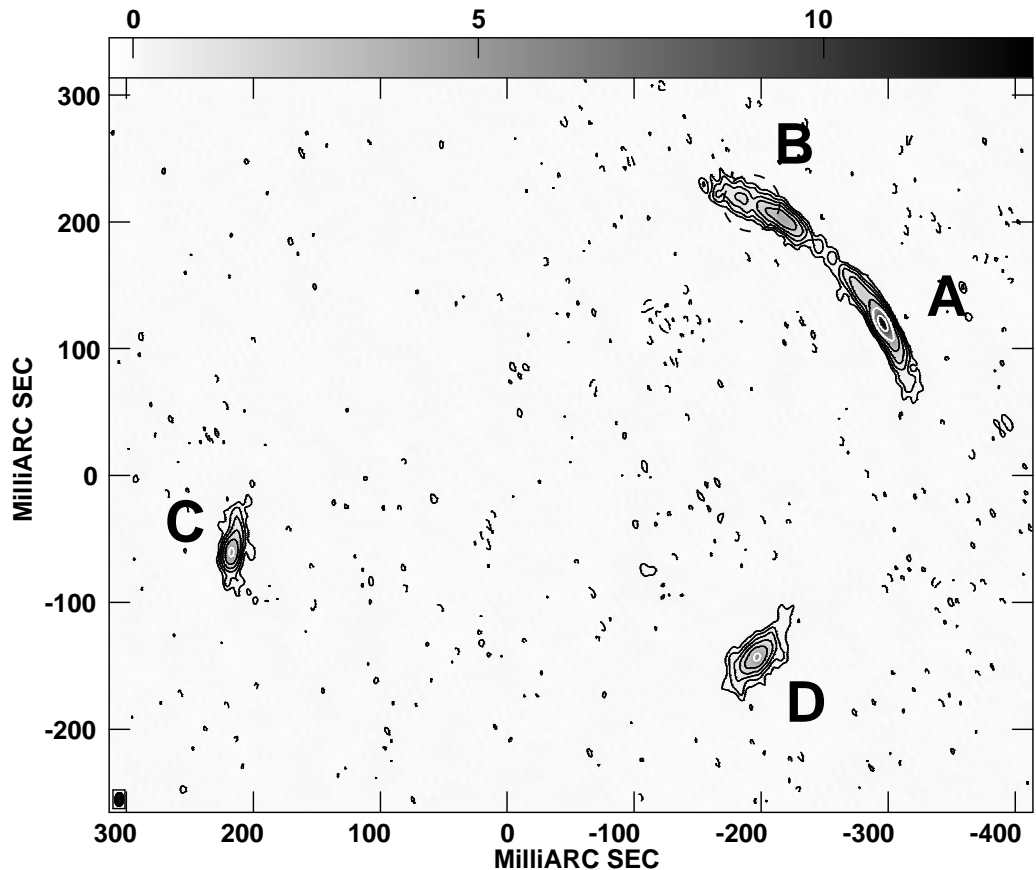}
\caption{EVN 21-cm map of CLASS B0128+437. The restoring beam is shown
in the bottom-left corner and has dimensions of $10.3\times5.9$~mas$^2$
at a position angle of $-3.3^{\circ}$. The greyscale gives surface
brightness in units of mJy\,beam$^{-1}$. A dashed circle shows the
area of scattering in image B.\label{evnmap}}
\end{figure*}

All calibration and mapping was carried out in {\sc aips}, the
combination of the low observing frequency and the relatively small
image separation enabling the whole system to be mapped in a single
``window''. The final self-calibrated map is shown in Fig.~\ref{evnmap}.

%

\section{Discussion}

\subsection{Scatter-broadening}

Prominent in the new map is
a ``hole'' in image B, a dashed circle marking its location
in Fig.~\ref{evnmap}. None of the other images contains a similar
feature, this being seen most easily in the brightest image, A. Being
the brightest, image A is also the largest and so if the hole was an
intrinsic feature of the lensed source it would have to be seen here
as well. The fact that the restoring beam is elliptical and aligned
approximately north-south i.e. with the jet, might cause the hole to
be somewhat obscured, but image A is peaked close to its centre whilst
image B is clearly brightest towards its western end. As the hole is only
seen in image B then its origin must be independent of the lensing
process and is, as already mentioned, due to scatter-broadening in
the ISM of the lens galaxy.

The position of the hole in image B is close to the location of
the 5-GHz structure seen in Fig.~\ref{vlbamaps} (the axis scales
in Figs~\ref{vlbamaps} and \ref{evnmap} are the same). Therefore,
either there is more scattering material in front of the 5-GHz
sub-components or these, being more compact than the extended
emission in the jet, are more obviously broadened. In fact, it
is probably a combination of these effects. There is certainly a
scattering gradient in image B as the eastern end of the jet is
much weaker than the western end and Fig.~\ref{vlbamaps} shows
that all the sub-components of an image should have comparable flux
densities. Also, model fitting to image A shows that the most compact
of the sub-components is the central one and so this should be most
affected by scattering. It seems likely though
that all of image B is subject to some level of scattering as
its surface brightness is consistently lower than that of image
A\footnote{Each image should have the same surface brightness.}
(two more contours are present in image A than in image B).

\begin{figure}
\centering
\includegraphics[scale=0.4]{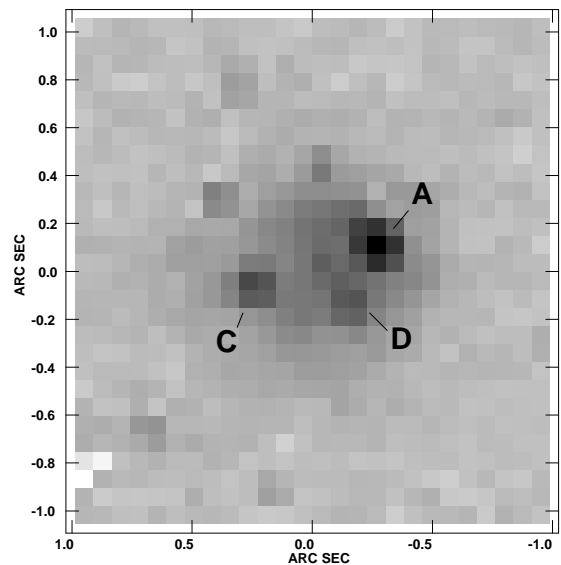}
\caption{NICMOS $H$-band image of B0128+437. Images A, C and D
have been marked.\label{nicmos}}
\end{figure}

Recent (February 2004) imaging with the NICMOS camera aboard the
Hubble Space Telescope (HST) support the hypothesis that a rich ISM
is responsible for distorting image B. Previous HST imaging in
$V$ and $I$ filters revealed that the system is extremely faint
in the optical (Biggs et al. \cite{biggs04}). Biggs et al. also showed
that in $K$-band, observed with the United Kingdom
Infrared Telescope (UKIRT), the system is much brighter
($K \sim 18$) although the resolution was too poor to separate the
lensed images. The new $H$-band HST image (Fig.~\ref{nicmos}) does
have sufficient resolution in the infrared to show the individual
images (and the lensing galaxy) although image B is conspicuous by its
absence. Dust obscuration is the most natural explanation and gas
associated with this dust would be responsible for the scatter-broadening.
The presence of large quantities of gas and dust, together with the small
separation of the lensed images, suggests that the lens galaxy is a
spiral.

An interesting hypothesis is that all the images are being
scatter-broadened and not just B. The reason for this is the
conflict between the observed structure of the images and the
radio spectrum of the lens system. Measurements of the
total flux density made at frequencies ranging from
325~MHz to 8.4~GHz show that the system has a Gigahertz peaked
spectrum (GPS), the spectrum exhibiting a pronounced turnover at
about 1~GHz. Given this, the source structure would be expected
to be compact, in contrast to the very extended nature of all
the images at 1.4~GHz. The alternative is that the source is not
actually a GPS, but that an intrinsic steep spectrum has been altered,
perhaps by free-free absorption, again in the ISM of the lensing
galaxy. A similar effect has been suggested in the lens JVAS
B2114+022 (Augusto et al. \cite{augusto01}). Helpful as a diagnostic
in this regard would be individual spectra of each image over a wide
range of frequency, a difficult task given the small separation of
the images. The NICMOS data, however, only provide clear evidence for
a rich ISM in front of image B and we consider the 4-image scattering
hypothesis unlikely.

\subsection{Lens mass modelling}

The growth of the image sizes with decreasing frequency, if not a
result of scatter-broadening, is excellent news for lens mass modelling
as the lens potential can be probed along a greater range of position
angles around the centre of the lens galaxy. Algorithms have been developed,
such as LensClean (Kochanek \& Narayan \cite{kochanek92}), which take as
input a radio map of a lens system and CLEAN this under the constraint
of a lens model. This has most recently been used on the
lens system JVAS B0218+357 (Wucknitz \cite{wucknitz04}). The result
of LensClean is a best-fit mass model, an image
of the unlensed source and an excellent map of the lens system
itself! Unfortunately, attempts to fit a model to the image
positions at 5~GHz in B0128+437 have been unsuccessful thus making
the choice of a starting model for LensClean difficult. The
scatter-broadening of image B also makes it more difficult to
LensClean the map although it is possible to add a scattering
coefficient into the computations.

Algorithms such as LensClean are most useful when a lens system
is dominated by extended structure. In those cases where the
images are resolved into discrete components, their positions
and sizes can be estimated with Gaussian modelfitting packages
(exemplified by that in {\sc difmap}) and these used as input
into standard lens modelling packages. As well as the extended
structure revealed in the new map two new discrete components
are seen in images A and B. These can be seen close to the
point where these images meet and constrain the position of
the ``critical curve'', a locus of points which determines where
images are created or destroyed.

\subsection{Substructure in the lens}

As mentioned in the previous section, the mass model for B0128+437 is
not currently well understood. Modelling presented in Biggs et al.
(\cite{biggs04}) suggests that the major problem is in fitting to
the position angle of the jet in image C (Fig.~\ref{vlbamaps}) which
lies along a position angle of approximately $-60^{\circ}$. A smooth
model such as a Singular Isothermal Sphere (SIE) with external shear
cannot reproduce the sub-component positions for this image, which
must mean that additional unmodelled substructure exists in the
lensing galaxy, perhaps that predicted by Cold Dark Matter (CDM)
models of large scale structure formation (e.g. De Lucia
\cite{delucia04}).

Given that image C is clearly extended in Fig.~\ref{evnmap} along
the north-south direction, the difference in the position angles
of the inner and outer jet might arise from mass substructure of
order $10^6$~M$_{\sun}$ rotating the inner jet, leaving the
larger-scale jet seen in lower frequency maps unperturbed. Models of
this system though do not seem to require a deviation from $-60^{\circ}$
of anything as extreme as that seen in the EVN map and a more plausible
explanation is that the extension at 1.4~GHz is simply the tangential
stretching of this image that is expected given that this is a
four-image system. The beginnings of this stretching can be seen
in Fig.~\ref{evnmap} as bulges to the jet towards the north and
south.

%

\section{Conclusions}

Of the 22 lens systems that were discovered during the course of the
JVAS/CLASS surveys, B0128+437 has proved to be one of the most
interesting, both for the distortion of image B and the inability of
smooth mass models to reproduce the observed image positions. Our new
EVN map shows the area of scatter-broadening in image B extremely
well and new NICMOS data confirm that a rich ISM obscures this image's
line of sight. An area of star formation, possibly contained within
a spiral arm, might be the origin of the dust and turbulent ionised
gas.

Future work on this system could include the following. High-frequency
VLBI (15~GHz) might be able to recover the position of the eastern
sub-component in image B, provided that the scatter-broadening had
sufficiently weakened by this point. The position would serve as a
useful model constraint. The other two sub-components have much steeper
spectra and would be almost certainly undetectable at the higher frequency.
On the other hand, low frequency VLBI
($<$1~GHz) could determine why the spectrum of the source turns over
at $\sim$1~GHz when the images are dominated by extended emission at
this frequency. Which parts of the images fade? Might an Einstein ring
appear? Finally, it might be
possible to detect redshifted HI absorption. The redshift of the lens
galaxy is not known although the detection of a single line with Keck
spectroscopy considerably narrows the search. Most possibilities have
drawn a blank using the Westerbork Synthesis Radio Telescope (WSRT), but
one redshift, $z=1.145$, has yet to be explored. If HI was found at this
redshift then its spatial distribution could be mapped with VLBI and
compared with the spatial variance of the scattering.

\begin{acknowledgements}

The European VLBI Network is a joint facility of European, Chinese, 
South African and other radio astronomy institutes funded by their 
national research councils. This work also included observations made
with the NASA/ESA Hubble Space Telescope, obtained at the Space
Telescope Science Institute, which is operated by AURA, Inc., under
NASA contract NAS 5-26555.

\end{acknowledgements}

\end{document}